\documentclass[14pt]{article}
\pdfoutput=1
\usepackage{amsmath,amssymb,amsfonts,amsthm,bm,bbm,cancel,wasysym}
\usepackage{epsfig,graphics,graphicx,epstopdf}
\usepackage{array,booktabs,colortbl,colordvi,multirow}
\usepackage{colordvi,color,xcolor}
\usepackage{hyperref}
\usepackage{rotating}

\usepackage{verbatim}
\usepackage{cite}
\usepackage{subfig}
\usepackage{setspace}
\usepackage{url}
\usepackage[percent]{overpic}
\usepackage{slashed}
\usepackage{authblk}
\usepackage{xspace}
\usepackage{fullpage}
\usepackage{hyperref}

\def\ie{{\it i.e.}}
\def\eg{{\it e.g.}}

\def\to{\rightarrow}

\newskip\zatskip \zatskip=0pt plus0pt minus0pt
\def\matth{\mathsurround=0pt}
\def\lsim{\mathrel{\mathpalette\atversim<}}
\def\gsim{\mathrel{\mathpalette\atversim>}}
\def\atversim#1#2{\lower0.7ex\vbox{\baselineskip\zatskip\lineskip\zatskip
  \lineskiplimit 0pt\ialign{$\matth#1\hfil##\hfil$\crcr#2\crcr\sim\crcr}}}

\begin{document}


\begin{flushright}
SLAC-PUB-17528\\
\today
\end{flushright}
\vspace*{5mm}

\renewcommand{\thefootnote}{\fnsymbol{footnote}}
\setcounter{footnote}{1}

\begin{center}

{\Large {\bf Dark Initial State Radiation and the Kinetic Mixing Portal}}\\

\vspace*{0.75cm}

{\bf Thomas G. Rizzo}~\footnote{rizzo@slac.stanford.edu}

\vspace{0.5cm}

{SLAC National Accelerator Laboratory}\\ 
{2575 Sand Hill Rd., Menlo Park, CA, 94025 USA}

\end{center}
\vspace{.5cm}

\begin{abstract}
 
\noindent

Data from Planck measurements of the cosmic microwave background (CMB) place important constraints on models with light dark matter (DM) and light mediators especially 
when both lie in the mass range below $\sim 1 $ GeV.  In models involving kinetic mixing where the dark photon acts as the mediator,  these constraints are easily satisfied and 
the appropriate DM relic density achievable if the DM is, \eg, a complex scalar, where $p$-wave annihilation occurs, or is the lighter component of a split pseudo-Dirac state where 
co-annihilation dominates. In both of these cases, although higher order in the dark gauge coupling, $g_D$, the corresponding annihilation processes including dark photon initial state 
radiation (ISR) will be dominantly $s$-wave with essentially temperature independent cross sections. The rates for these dark ISR associated processes, though not yielding cross 
sections large enough to contribute to the relic density,  can still run into possible conflicts with the bounds arising from the CMB. In this paper we perform a preliminary study of the present 
and potential future constraints that the CMB imposes on the parameter spaces for both of these scenarios due to the existence of this dark ISR. Further analyses of the effects of 
dark ISR in DM annihilation is clearly warranted.
\end{abstract}

\renewcommand{\thefootnote}{\arabic{footnote}}
\setcounter{footnote}{0}
\thispagestyle{empty}
\vfill
\newpage
\setcounter{page}{1}



\section{Introduction}

The nature of dark Matter (DM) and its possible interactions with the particles of the Standard Model (SM) other than via gravity remains a great mystery. While Weakly Interacting Massive 
Particles (WIMPs)\cite{Arcadi:2017kky,Roszkowski:2017nbc} and axions\cite{Kawasaki:2013ae,Graham:2015ouw,Irastorza:2018dyq} remain as quite viable contenders for the role 
of DM, increasingly sensitive experiments have failed to yield any convincing signals\cite{LHC,Aprile:2018dbl,Fermi-LAT:2016uux}  
for these long anticipated states.  The continued shrinking of the allowed parameter spaces for these scenarios has stimulated a rapid growth of new DM models covering 
exceedingly wide ranges in both possible masses and couplings\cite{Alexander:2016aln,Battaglieri:2017aum,Bertone:2018krk,Amole:2019fdf}. 
As current experiments seemingly disfavor the interaction of DM with 
us via SM strength couplings, other new interactions must likely exist to explain, \eg, how DM reaches its observed relic abundance\cite{Steigman:2015hda,Saikawa:2020swg}. One way 
to classify such new interactions is via a set of `portals' linking the SM with 
fields in the dark sector. Only a few renormalizable, dimension-4 portals exists; of these, the vector boson/kinetic mixing (KM) portal has received a wide amount of attention in the 
literature\cite{KM,vectorportal} and will be the subject of our discussion below. In the simplest version of such a scenario, the DM fields are SM singlets but are instead charged under a 
new $U(1)_D$ gauge interaction, with corresponding gauge coupling $g_D$,  mediated by a dark photon (DP) \cite{Fabbrichesi:2020wbt} whose mass can be generated by the dark 
analog of the SM Higgs mechanism. This DP 
then experiences KM with the SM $U(1)_Y$ hypercharge gauge boson at, \eg, the 1-loop level via a set of `portal matter' fields which are charged under both gauge 
groups\cite{Rizzo:2018vlb,Rueter:2019wdf,Kim:2019oyh}. After the gauge fields are canonically normalized and the usual SM and dark spontaneous symmetry breakings occur, 
the DP picks up a small coupling to the SM fields. For a DP in the mass range below a 
few GeV, this coupling is quite well approximated simply by $\simeq \epsilon eQ_{em}$, where $\epsilon$ is a small dimensionless parameter, $\sim 10^{-(3-4)}$ in the present discussion, 
that describes the magnitude of this loop-suppressed KM. 

The KM scenario is of special interest when the DM and DP are both relatively light $\lsim 1$ GeV as in such a case the DM can be a thermal relic in a manner similar to what occurs in 
the WIMP scenario{\footnote {The DP and DM masses are naturally similar in several scenarios, \eg, models with extra dimensions where the scale of their masses is set by the (inverse) 
size of the  compactification scale, $R$. See, \eg, Refs.~\cite{Rizzo:2020ybl,Landim:2019epv,Rizzo:2018joy,Rizzo:2018ntg}.}. 
For such a range of masses, pair annihilation of DM to achieve the proper relic density via the exchange of a virtual DP usually results in pairs 
of electrons, muons, or light charged hadrons. For DM at the $\sim 10-1000$ MeV mass scale{\footnote {The rough lower bound on the DM mass of $\sim 10$ MeV is taken from 
Ref.\cite{Sabti:2019mhn}}}, constraints from Planck\cite{Aghanim:2018eyx} on the CMB tell us that at $z\sim 10^3$ 
the DM annihilation cross section into light charged states, \eg, $e^+e^-$, must be relatively suppressed\cite{Slatyer:2015jla,Liu:2016cnk,Leane:2018kjk} to avoid injecting additional 
electromagnetic energy into the plasma{\footnote {Important constraints of a very similar magnitude also arise from data from Voyager 2\cite{Boudaud:2016mos,Boudaud:2018oya}.}. 
This constraint lies roughly\cite{Cang:2020exa} at the level of  $\sim 5 \times 10^{-29}~(m_{DM}/100~ {\rm {MeV}})$ cm$^3$s$^{-1}$, and is seen to depend approximately linearly on the DM 
mass so that it becomes relatively inconsequential above masses $\gsim 10-30$ GeV. Further it is to be noted that this constraint is expected to improve by roughly a factor of $\sim 2-3$ in 
the coming years\cite{Green:2018pmd,Ade:2018sbj,Abazajian:2016yjj}. 
This bound for DM annihilation in this $\sim 10-1000$ MeV mass range, however,  lies several orders of magnitude 
below the cross section required to reach the observed relic density during freeze out, \eg, $<\sigma v_{rel}>_{FO} \simeq 4.4(7.5)\times 10^{-26}$ cm$^3$s$^{-1}$ for an 
$s$-wave annihilating, self-conjugate fermion (or a $p$-wave annihilating complex scalar) DM\cite{Steigman:2015hda,Saikawa:2020swg}. 
This requirement puts a strong constraint on the nature of the DM interacting with the SM via the DP as well as on their relative masses. For example, if 
$m_{DM}>m_{DP}$, then DM can annihilate into a pair of DPs which will likely be the dominant mechanism to achieve the observed relic density. As an $s$-wave process, the cross 
section for this reaction during freeze out and during the time of the CMB would naturally be quite similar and so this reaction, as well as those for other $s$-wave annihilation processes, 
would necessarily be forbidden for light DM by the above mentioned constraints. Clearly a set of mechanisms that reduce the DM annihilation cross section 
as the temperature, $T$, decreases are needed for DM and DP in this mass range. 

One obvious way to avoid this constraint is to require that $m_{DM}< m_{DP}$ and also that the $s$-channel DP-mediated annihilation process be $p$-wave so that it is 
velocity-squared suppressed at times much later than freeze out;  this is quite helpful since $v_{rel}^2 \sim T$. For spin-1 mediators such as the DP,
this possibility restricts the DM to be a Majorana fermion{\footnote {The Majorana fermion DM has only have axial-vector couplings to the DP while the DP simultaneously would have  
vector-like couplings to the SM. This  is not realized within the present setup without more complexity in the dark sector, \eg, additional fermions to cancel gauge anomalies.}} 
or a complex scalar, $\phi$. Note that since Dirac fermion DM annihilation via an $s$-channel, 
spin-1 exchange with only vector couplings is dominantly $s$-wave, this possibility is forbidden within the above range of masses. A second scenario is that the DM is pseudo-Dirac, forming 
two mass eigenstates, $\chi_{1,2}$, that are split in mass and which co-annihilate to the SM via the DP. For a fixed mass splitting, as the temperature drops this co-annihilation 
process becomes highly Boltzmann suppressed thus avoiding the CMB constraints. The parameter spaces of these two classes of models for light DM/DP in the mass range of interest 
to us here have been widely studied\cite{Alexander:2016aln,Battaglieri:2017aum} and it is known that the observed relic density is achievable in both of 
these setups while still satisfying all other existing experimental constraints. 

For SM singlet DM, the lifting of velocity (and/or helicity) suppression of DM annihilation in many models via the final state radiation (FSR) of SM gauge bosons off of other SM fields in the 
final state is well-known\cite{Bell:2008ey,Kachelriess:2009zy,Bringmann:2012vr}. In our mass range of interest, the only possible on-shell FSR would be via photons being radiated off 
of, \eg, an $e^+e^-, ~\mu^+\mu^-$ or light charged hadron final state. 
Less well studied, but of similar, and in some cases more, importance is the initial state radiation (ISR) of the dark mediator field, here the DP, off of the annihilating DM\cite{Bell:2017irk}. 
Note that since the DM can only be a SM singlet in the present scenario, such ISR {\it {must}} necessarily be dark at tree-level. Given the discussion above, this process can obviously 
be of most relevance to us when $m_{DM}<m_{DP}<2m_{DM}$ which is a relatively narrow mass window but is of important experimental interest since within it any DPs produced 
on-shell can decay only to SM final states. There are many existing and proposed searches for such classes 
of events\cite{Adrian:2018scb,CasaisVidal:2019nnb,Ariga:2019ufm,Banerjee:2019hmi,Lubatti:2019vkf,Campajola:2019nls,Ahdida:2020evc,Akesson:2018vlm}.  Fig.~\ref{fig12}, for the 
case of complex scalar DM, shows the usual annihilation mechanism via the $s$-channel exchange of a DP in the left panel while the analogous process with additional dark ISR is shown 
symbolically in the right panel. Note that in the specific case of complex scalar DM, the 4-point process also contributes besides the $t$- and $u$-channel graphs while such a contribution 
is absent for the case of pseudo-Dirac DM.

\begin{figure}[htbp]
\centerline{\includegraphics[width=2.5in]{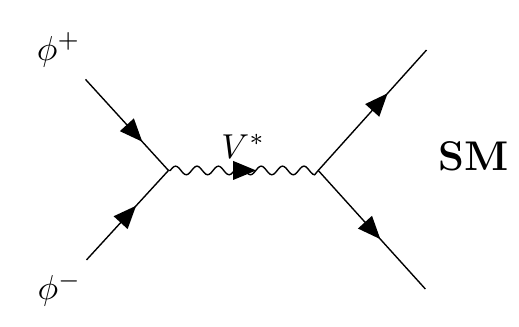}
\hspace{1.1cm}
\includegraphics[width=3.2in]{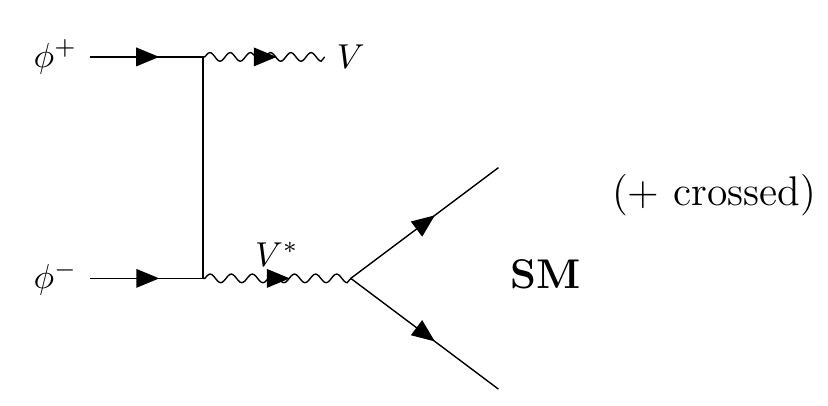}}
\vspace*{-0.20cm}
\caption{The two basic processes of interest to us in this paper symbolically employing the case of complex scalar DM as a representative example. (Left) DM annihilation into SM final 
states via $s$-channel DP exchange. (Right) The same process with an additional DP emitted from the initial state producing $u-$ and $t-$channel graphs. For the case of complex 
scalars the 4-point vertex graph (not shown) is also present. }
\label{fig12}
\end{figure}

We note, and as will be examined further below, that dark ISR can, in principle, lead to some affects which 
are not possible via FSR: ($i$) In the case of complex scalar DM, FSR cannot lift the $p$-wave suppression since this is the result of the (non-relativistic) coupling of the two spin-0, 
on-shell initial states with the spin-1, off-shell DP. This is easy to see as in the non-relativistic limit, the usual scalar coupling, $(p_1-p_2)_\mu$ has, to $O(v_{rel}^2)$, only non-zero 
3-spatial components, $\simeq m_{DM}v_{rel}$, so that this automatically leads to a $p-$wave process. Dark ISR changes this situation in that now one of the DM particles in the initial 
state annihilating via the virtual DP, itself goes off-shell. We now make a rather simple observation: although this process is both $\sim g_D^2/4\pi =\alpha_D $ and 3-body phase 
space suppressed, it is still 
dominantly an $s$-wave annihilation mechanism, though it is too small to make any significant contribution to the DM relic density. However,  it can yield a rate which is sufficiently large 
so as to be in potential conflict with the CMB constraints above and thus lead to restrictions on the model parameter space; we will examine this possibility below. ($ii$) Similarly, in 
the case of pseudo-Dirac DM, the two DM mass eigenstates couple in an off-diagonal manner to the DP while the `diagonal' couplings, which would be unsuppressed $s$-wave 
annihilations, are absent at lowest order in $g_D$. It is well known that obtaining the observed relic density via the co-annihilation process requires that the relative mass splitting of these 
two eigenstates be relatively small, \ie, $\delta =\Delta m/m_{DM} <<1$. If not, this process becomes highly Boltzmann suppressed by a factor of order $\sim e^{-\delta x_f}$ with 
$x_f=m_{DM}/T_{FO} \simeq 20-30$. As we will see below, dark ISR may partially negate this (very) large Boltzmann suppression even in cases where this mass splitting is no longer 
so small, \ie,  $\delta \lsim 0.25$, but at the cost of this additional power of $\sim \alpha_D$ and 3-body phase space. The resulting rate is, 
however, too small to make a significant contribution to the relic density. But in this 
case too, an $s$-wave process is the net result so that the question again arises as to whether or not the bounds from the CMB remain satisfied once ISR of a DP occurs. We will 
address this issue within this model context in the subsequent analysis as well. Clearly the implications of ISR dark radiation for the allowed parameter space of a given DM model 
warrants some further examination and that is goal of the preliminary analysis below.

The outline of this paper is as follows: in Section 2, we will discuss the general overarching model structure that we will consider and then immediately turn our attention to the specific 
case of complex scalar 
DM. In particular, we will explore the extent to which dark ISR can lead to conflicts with the constraints arising from the CMB and the corresponding restrictions this imposes on the 
complex DM model parameter space. Similarly, in  Section 3, we will explore the impact of dark ISR in the case of pseudo-Dirac DM wherein co-annihilation is the dominant process leading 
to the observed relic density. We again will examine the possible conflict with the CMB constraints that can arise and which lead to restrictions on the model parameter space even in cases 
where the mass splitting between the two eigenstates is significant and contrast this with the previously examined case of complex scalar DM.  A discussion and our conclusions 
can be found in Section 4.



\section{General Model Features and the Complex Scalar DM Scenario}

We first briefly summarize the common features of the basic framework that we consider below, restate our essential assumptions and then establish subsequent notation before 
continuing with our analysis.

\subsection{Basic Setup}

The interactions in the (mostly) dark gauge/Higgs sector of our model in the original weak eigenstate basis are described by the general Lagrangian
\begin{equation}
L =-\frac{1}{4} \hat V_{\mu \nu} \hat V^{\mu \nu} -\frac{1}{4} \hat B_{\mu\nu} \hat B^{\mu\nu} 
+\frac{\epsilon}{2c_w} \hat V_{\mu\nu} \hat B^{\mu\nu} + (D_\mu S)^\dagger (D^\mu S) -U(S^\dagger S)- \lambda_{HS} H^\dagger H S^\dagger S+L_{SM}  \,,  
\end{equation}
where $\hat V, \hat B$ are the kinetically mixed $U(1)_D$ DP and the SM weak $U(1)_Y$ hypercharge gauge fields, respectively, with the strength of this KM being described by the  
dimensionless parameter $\epsilon$; here $c_w=\cos \theta_w$ with $\theta_w$ being the usual SM weak mixing angle. Since we will assume that the KM parameter is very small in what 
follows, $\epsilon \simeq 10^{-(3-4)}$, except where necessary we can work to leading order in $\epsilon$. In this limit the KM is removed by the simplified field redefinitions 
$\hat B \to B+\frac{\epsilon}{c_w}V,~\hat V \to V$. $H,S$ denote the SM and dark Higgs fields, respectively, while $L_{SM}$ describes the rest of the interactions of the SM. The 
vacuum expectation value of $S$, $v_s/\sqrt 2$, resulting from minimizing 
the potential $U$, generates the mass of the DP, $m_{DP}=m_V=g_DQ_D(S)v_s$, via the dark covariant derivative $D_\mu=\partial_\mu+ig_D Q_D(S)\hat V_\mu$.  Here, $g_D$ is the 
$U(1)_D$ gauge coupling and $Q_D(S)$ is the relevant dark charge of $S$. Due to the vevs of both $H$ and $S$, $\lambda_{HS} \neq 0$ generates a mass 
mixing between the SM and dark Higgs fields which we will necessarily assume to be very small to avoid the strong bounds arising 
from invisible Higgs decays\cite{ATLAS:2020cjb} and so will be considered to be phenomenologically irrelevant in the subsequent discussion. It will also be assumed that $m_S>(1-2)m_V$, similar to what happens in the SM, so that $S$ is guaranteed to be unstable and can decay sufficiently  
rapidly as, \eg, $S\to 2V$ (or $\to VV^*$ depending on the mass ratio) although the specific details of this will not be required for the discussion below.

\subsection{Complex Scalar DM Scenario} 

While the form of $L$ above is quite general, the addition of the DM itself will introduce some new terms to this set of interactions. In this subsection, we consider the case where the 
DM is a complex scalar, $\phi$; the form of the general additional pieces of the Lagrangian in this case are given by 
\begin{equation}
L_{DM}  =(D_\mu \phi)^\dagger (D^\mu \phi) -U_\phi(\phi^\dagger \phi)-( \lambda_{H\phi} H^\dagger H+\lambda_{\phi S} S^\dagger S)  \phi^\dagger \phi\,,  
\end{equation}
where $D_\mu$ is the relevant covariant derivative for $\phi$, whose dark charge will be taken to be unity in what follows without loss of generality.  The potential, $U_\phi$, describes 
the (taken to be weak) DM self-interactions which we assume do not generate a vev for $\phi$ so that it can remain stable. The quartic couplings of $\phi$ with both $S$ and $H$, 
described by the coefficients $\lambda_{\phi (S,H)}$, respectively, will also be assumed to be quite small and to play no essential role in what follows. 

Under this set of assumptions, the 
DM-DP gauge interaction completely dominates and leads to a slightly modified version of the familiar expression from, \eg, Ref.\cite{Berlin:2014tja} for the annihilation cross section of scalar 
DM pairs into SM fields via the process in the left panel of Fig.~\ref{fig12} in our mass range of interest; summing over electron, muon and light charged hadronic final states, this is given by 
\begin{equation}
\sigma_{DM} v_{rel}=\frac{g_D^2\epsilon^2 e^2}{6\pi}~\frac{s\beta_\phi^2}{(s-m_V^2)^2+(\Gamma_Vm_V)^2} ~\Big[1+\frac{\beta_\mu(3-\beta_\mu^2)}{2}~\theta(s-4m_\mu^2) +R~\theta(s-4m_\pi^2)\Big]\,,  
\end{equation}
where $s$ is the usual center of mass energy,  $\beta_{\phi,\mu}^2=1-4m_{\phi,\mu}^2/s$ are the squares of the DM (muon) velocities, $\Gamma_V$ is the total decay width of the DP, 
and $R$ is the familiar cross section ratio  $R=\sigma(e^+e^- \to hadrons)/\sigma(e^+e^-\to \mu^+\mu^-)$ via virtual photon exchange for massless muons. Of course, this hadronic 
contribution only turns on above the two-pion threshold, $\sim 280$ MeV; note that the electron mass 
has been neglected in the expression for $\sigma_{DM}v_{rel}$. Here we observe the explicit $\beta_\phi^2$-dependence of the DM annihilation cross section as expected 
from the discussion above.  Note that $\Gamma_V$ depends sensitively on the value of $r=m_\phi/m_V$ since for $r<1/2$, the DP can decay dominantly into DM pairs with a 
reasonable large partial width,  $\Gamma_V/m_V =g_D^2 \beta^3/48\pi~${\footnote {Here it is assumed that $g_D$ is very roughly the size of a typical 
gauge coupling.}}, where $\beta^2=1-4r^2$. However, for larger values of $r$, in the kinematic region of interest to us below, only the DP decay to SM states is 
allowed and these modes all have partial widths that are $\epsilon^2$ suppressed, \eg, $\Gamma (V \to e^+e^-)/m_V=(\epsilon e)^2/12\pi \lsim 10^{-(8-10)}$, where the electron mass has 
again been neglected.  

\begin{figure}[htbp]
\vspace*{-0.5cm}
\centerline{\includegraphics[width=5.0in]{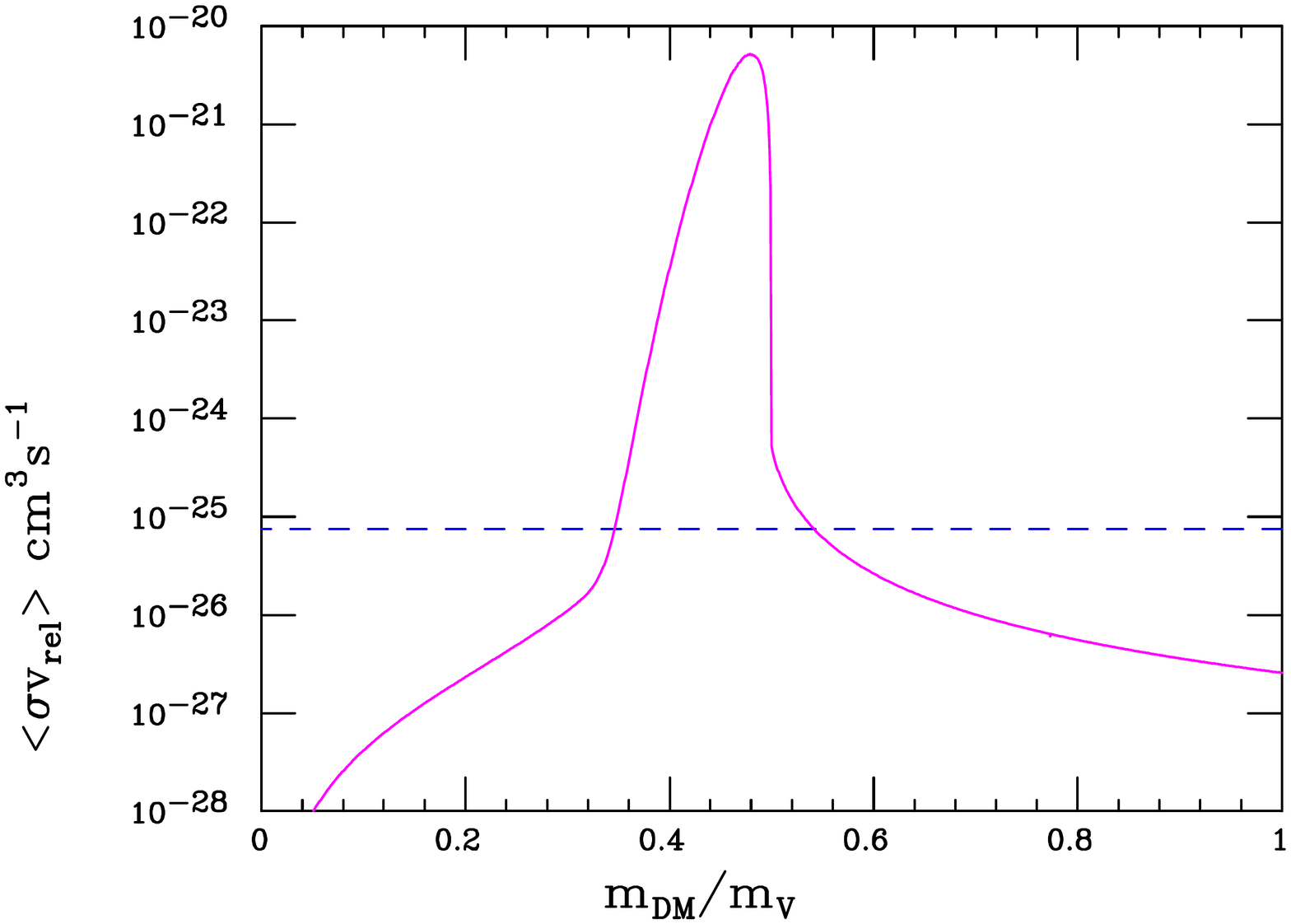}}
\vspace*{-2.2cm}
\centerline{\includegraphics[width=5.0in]{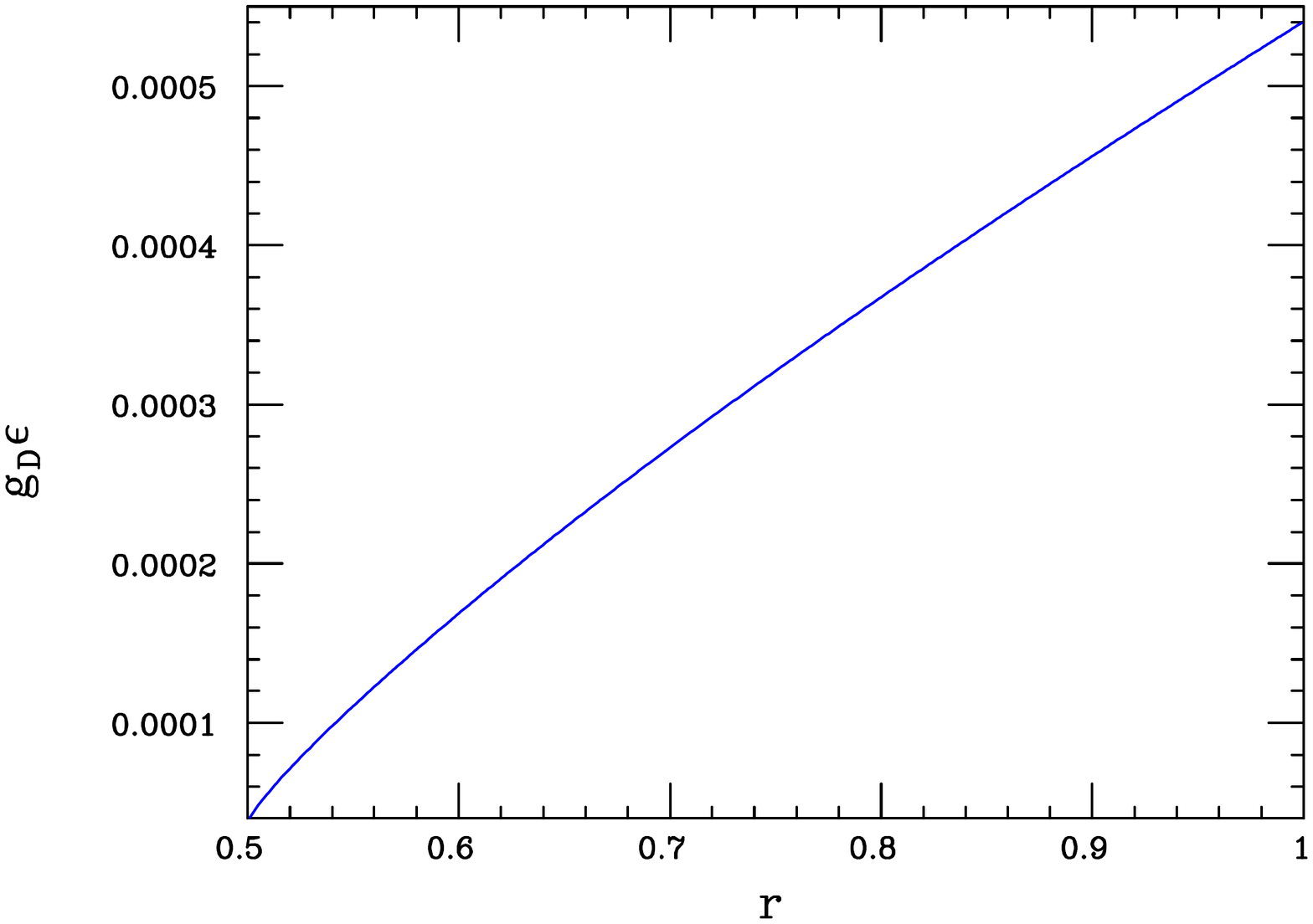}}
\vspace*{-1.50cm}
\caption{(Top) Sample thermally averaged complex scalar DM annihilation cross section as a function of $r=m_\phi/m_V$, assuming $m_V=100$ MeV and $g_D\epsilon=10^{-4}$ with 
$x_f=20$. The expression in the square 
bracket in the text has thus been set to unity here. The corresponding results for other values of these input parameters can be obtained via simple rescaling using the equation in the text. 
The dashed line shows the approximate annihilation cross section needed to obtain the observed relic density and thus for any $r=m_{DM}/m_V$ in our range of interest 
the required value of $g_D\epsilon$ can be determined. (Bottom) The value of $g_D\epsilon$ required to obtain the observed relic density as a function of $r$ in our range of interest 
assuming that $m_V=100$ MeV based on the results presented in the top panel.}
\label{fig1}
\end{figure}

To obtain the thermally averaged cross section leading to the DM relic density we must integrate the above expression weighted by the Bose-Einstein distributions 
of the incoming complex DM states. Though this is $p$-wave suppressed, the thermal average of the above cross section can yield the observed relic density for the $1/2<r<1$ kinematic 
range of interest as is shown in Fig.~\ref{fig1}. Special care must be taken with the resonance enhancement region\cite{Feng:2017drg,Li:2015tka} which we take some minor 
advantage of in parts of the phase space. For example, if one assumes $g_D\epsilon =1 \times 10^{-4}$ and $m_V =100$ MeV, as employed for purposes of demonstration 
in top panel of Fig.~\ref{fig1},  the required annihilation cross 
section is obtained near $r\sim 0.55$, within our mass range of interest. Similar other combinations of parameter choices within the mass range $1/2<r<1$ can work equally well.  As 
seen in the lower panel of Fig~\ref{fig1}, fixing $m_V=100$ MeV for purposed of illustration, we observe that for each value of $r$ there is a unique value of the product $g_D\epsilon$ 
which yields the observed relict density. Then, for any given value of $\epsilon$, the value of $g_D$ is completely determined as a function of $r$ and must increase as $r$ 
increases. This 
implies that the larger $r$ region will be more sensitive to dark ISR effects since these scale as $g_D^4$. We will require that value of $g_D$ determined in this manner remain perturbative 
in the analysis below, \ie, $g_D \leq \sqrt {4\pi}\simeq 3.55$; this is relevant as decreases in $\epsilon$ must be compensated for by corresponding increases in the value of $g_D$. 
Since $\sigma_{DM} v_{rel} \sim (g_D\epsilon/m_V)^2$, increasing the value of $m_V$ for any fixed value of $r$ will necessitate a similar increase in $g_D\epsilon$, apart from the effect 
of new SM final state mass thresholds, in order to achieve the observed relic density. Away from hadronic resonances the effect of new thresholds is not too large, \eg, for $m_V=500$ 
MeV, the total annihilation 
cross section is found to be roughly $\simeq 3.5$ times larger that that given by just the $e^+e^-$ mode alone due to the additional contributions arising from muons and pions. Such 
effects can be easily accounted for in determining the desired values of $g_D\epsilon$ to maintain the relic density prediction. 

Clearly, we also observe that other equally valid solutions will exist when $r<1/2$ but they lie outside the specific mass region of interest to us here. Of course, this overall parameter space 
will be further constrained by other future experiments, \eg, via dark photon production at accelerators or via DM direct detection experiments employing $\phi$ scattering off of 
electrons and/or nucleons. 

We turn now to our first example of the influence of dark ISR. Consider the specific process $\phi^\dagger \phi \to VV^*, V^*\to e^+e^-$, which occurs via both $t-$ and $u-$channel 
DM exchanges as well as via the usual 4-point coupling as noted in Fig.~\ref{fig12}.  We will consider this reaction in the non-relativistic limit for the DM in the mass range of interest 
$1/2<r <1$; we will 
continue to take $m_e=0$ for simplicity in what follows as we will always assume that $m_{DM}>>m_e$. We note that if $\lambda_{\phi S}$, discussed above, were significant, something 
that we have assumed {\it {not}} to be the case here, then an additional $s$-channel diagram would be possible via virtual $S^*$ exchange followed by $S^*\to VV^*$. We find, however, 
that even when $\lambda_{\phi S}$ is significant, this contribution would be relatively numerically suppressed (by up to one or two orders 
of magnitude) due to a combination of overall constant factors combined with a relative ratio of $(m_V/m_S)^4 <<1$ in comparison to that coming from DP exchange. In any case, 
we will ignore this possible contribution here.

Recall that these $t-,u-$channel DM exchanges added to the 4-point interaction result in an $s$-wave process and so it is subject to the above mentioned constraints from the CMB. 
We can write the numerical result for the annihilation cross section for this reaction as 
\begin{equation}
\sigma_{DM} v_{rel} =0.216~\sigma_0 ~g_D^2 ~\Big(\frac{g_D\epsilon}{10^{-4}}\Big)^2 ~\Big(\frac{100 \rm {MeV}}{m_V}\Big)^2~I\,,  
\end{equation}
where $\sigma_0=10^{-26}$ cm$^3$s$^{-1}$ sets the scale to that which is roughly required for the annihilation cross section to result in the observed DM relic density and $I$ is the 
3-body `phase space' integral 
\begin{equation}
I=2q^2\int_0^{(1-q)^2} dx_{12}\int_{min}^{max} dx_{23}~\frac{2(1-x_{23})(x_{12}+x_{23}-q^2)-x_{12}(1+q^2+x_{12})}{(1-x_{12}-q^2)^2 [(x_{12}-q^2)^2+(Gq^2)^2]}\,,  
\end{equation}
where $q=1/(2r)=m_V/2m_\phi$, $G=\Gamma_V/m_V$, $x_{12}=m_{ee}^2/4m_\phi^2$, with $m_{ee}$ being the $e^+e^-$ invariant mass in the final state,  $x_{23}=1-E_{e^-}/m_\phi$ 
with $E_{e^-}$ being the corresponding electron energy, and the range of the $x_{23}$ integration is given by 
$(max,min)=\frac{1}{2}(1-x_{12}+q^2)\pm \frac{1}{2}[(1-x_{12}-q^2)^2-4x_{12}q^2]^{1/2}$. In terms of the parameter $q$ appearing in the integral, 
we note that the $\phi^\dagger \phi \to V^* \to e^+e^-$ resonance 
region lies in the vicinity of $q\simeq 1$ while the on-shell $\phi^\dagger \phi \to 2V$ process occurs when $q\sim 1/2$. Since the reaction with ISR is an $s$-wave process, the 
thermal average cross section (sufficiently far from the resonance) is essentially the same as the non-relativistic cross section itself, \ie, 
$<\sigma_{DM} v_{rel}> \simeq \sigma_{DM} v_{rel}$, so that numerical 
results can be obtained in a rather straightforward manner employing the leading term in the familiar velocity expansion. 
This will be a very good approximation at the the time of the CMB where the DM is 
quite non-relativistic. Since the DM kinetic decoupling temperature is $\sim 10^{-3} m_{DM}$\cite{Bringmann:2006mu},  at CMB times the DM temperature, $T_{DM}(CMB)$, can be 
roughly estimated as 
\begin{equation}
T_{DM}(CMB) \lsim 10^{-3} m_{DM}\Big(\frac{T_{CMB}}{10^{-3}m_{DM}}\Big)^2 <<m_{DM}\,,  
\end{equation}
Thus at the time of the CMB in which we are interested the higher order terms in $v_{rel}^2\sim T_{DM}(CMB)$ will necessarily be quite highly suppressed and totally ignorable

\begin{figure}[htbp]
\centerline{\includegraphics[width=5.0in,angle=0]{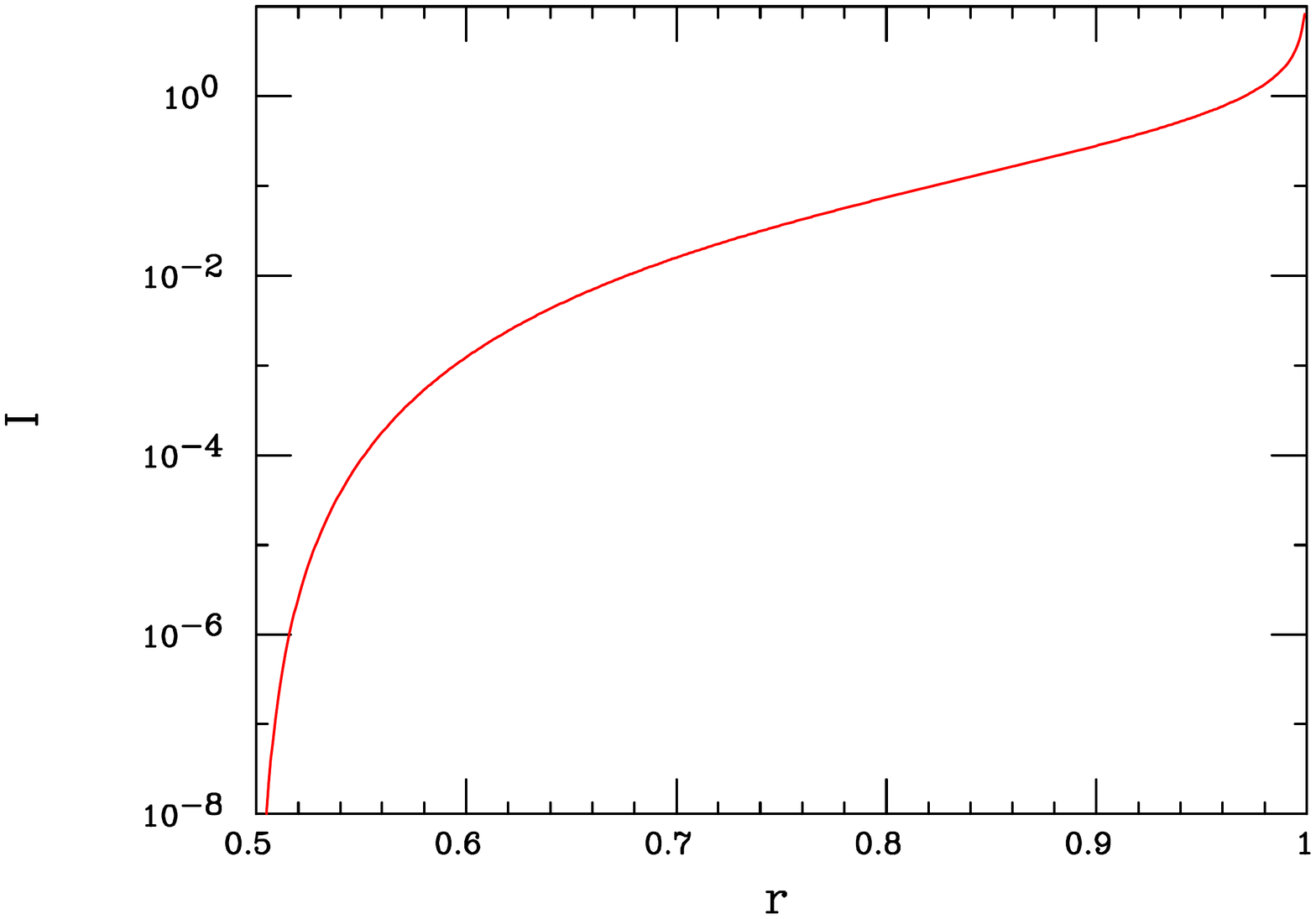}}
\vspace*{-2.2cm}
\centerline{\includegraphics[width=5.0in,angle=0]{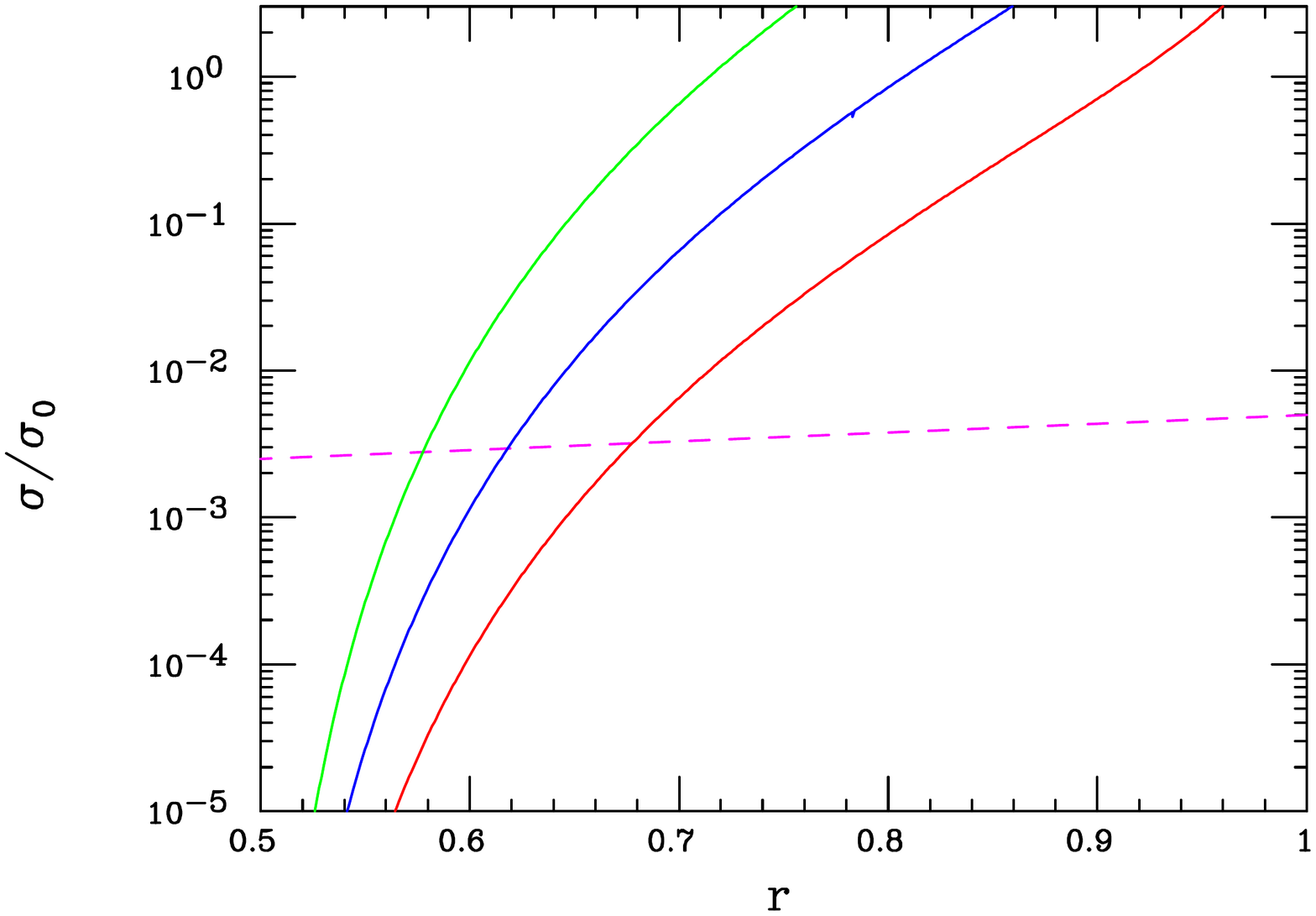}}
\vspace*{-1.30cm}
\caption{ (Top) Value of the phase space integral $I$ as defined in the text as a function of the parameter $r$. Recall that as $r\to 1$ both DPs can become on-shell. 
(Bottom) The cross section for the $s$-channel process 
$\phi^\dagger \phi \to VV^*, V^*\to e^+e^-$ with $v_{rel}^2=0$, in units of $\sigma_0$,  as a function of $r$ assuming, from bottom to top, that $\epsilon=10^{-(3,3.5,4)}$ and assuming that 
$m_V=100$ MeV. The approximate upper bound on this cross section arising from the CMB\cite{Cang:2020exa} is shown as the dashed line.}
\label{fig2}
\end{figure}

Our approach will now be as follows: taking $m_V=100$ MeV for purposes of demonstration, we will input into the above cross section 
expression the values of $g_D\epsilon$ that are required to obtain the observed relic density as a function of $r$ as displayed above in the lower panel of Fig.~\ref{fig1}. Next, we will 
assume as further input three representative, phenomenologically interesting values for $\epsilon=10^{-3},~10^{-3.5}$ and $10^{-4}$, respectively, from which $g_D$ can then be obtained 
as a function of $r$, subject to the perturbativity constraint mentioned above, \ie, $g_D\leq \sqrt {4\pi} \simeq 3.55$.  We note that in some cases, that if $r$ is sufficiently large this 
bound can be violated so some care is required. Note that with the relic density constraint used as input the dark ISR cross section has a practical scaling $\sim \epsilon^{-2}$ which 
will make the c ross section larger and hence the CMB bound {\it stronger} as $\epsilon$ decreases.  Finally, 
as is well known, the 3-body phase space integral $I$ will itself have a rather strong dependence on the value of $r$ as is shown in the upper panel of Fig.~\ref{fig2}. Combining these 
factors we then obtain the resulting cross section for the $\phi^\dagger \phi \to VV^*, V^*\to e^+e^-$ process as a function of $r$ as shown in the lower panel 
of Fig~\ref{fig2} where we also compare it to the approximate upper limit obtained from the CMB in Ref.\cite{Cang:2020exa}.  Note that along each curve of constant $\epsilon$, $g_D$ 
is increasing from left to right as $r$ increases to maintain the appropriate DM relic density and its value can be extracted directly from the lower panel of Fig.~\ref{fig1}. 

Several things can be seen from this Figure: ($i$)  As we might have expected, as $\epsilon$ decreases the value of $g_D$ increases for fixed $m_V$ to maintain the correct relic 
density and this results in an increase in the 3-body cross section for a given value or $r$. ($ii$) In particular, for $\epsilon=10^{-(3,3.5,4)}$, to satisfy the approximate CMB bound we must 
have $r\lsim (0.68,0.62,0.58)$, respectively; 
larger values of $r$ in each case are thus {\it excluded}.  These are respectably strong constraints on the model parameter space and as $\epsilon$ becomes smaller 
we are pushed ever closer to the DP resonance region for DM annihilation to maintain smaller values of $g_D$.  As noted earlier, since the 3-body 
cross section effectively scales as $\sim \epsilon^{-2}$ when the relic density is used as a constraint the results for other values of this parameter are easily obtained. ($iii$) This $r$ bound 
then results in the further constraint that $g_D\lsim (0.28, 0.68, 1.67)$, respectively,  
for these chosen values of $\epsilon$, all of which lie within the perturbatively allowed range. ($iv$) If the value of $m_V$ increases for fixed values of $r$ and $\epsilon$, then $g_D$ must also increase to maintain agreement with the observed relic density. This implies that the cross section 
for the 3-body 
$\phi^\dagger \phi \to VV^*, V^*\to e^+e^-$ process must also grow but now even more rapidly since it scales as $g_D^4$. For example, a doubling of $m_V$ to 200 MeV for fixed values of 
($r,\epsilon$), quadruples the 3-body annihilation rate. However, at the same time, the approximate CMB cross section constraint is also weakened, \ie, increased by a factor of two so 
that the ratio of the predicted 3-body cross dark ISR section to the CMB bound only doubles. Still, due to the steep rise of the cross section with increasing $r$, the resulting bound on $r$ 
strengthens only slightly by roughly $\simeq 0.02$. Further increases in the value of $m_V$ will thus only somewhat tighten the constraint on the value of $r$ obtained above. We note that 
future constraints on this 3-body process from the CMB are only expected to strengthen by roughly at most a factor of $\sim 2-3$ in the coming few 
years\cite{Green:2018pmd,Ade:2018sbj,Abazajian:2016yjj} so that given the rapid growth of the 3-body cross section with $r$ the resulting improved restrictions on the parameter space 
of this model will be rather minor but perhaps not ignorable in a detailed numerical analysis. A final thing that we should notice in the Figure is that for large values of $r$ the 3-body 
$\phi^\dagger \phi \to VV^*, V^*\to e^+e^-$ process could have made a genuine contribution to the DM relic density if this region had not been disallowed by the CMB data.

It is clear from this analysis that the CMB can place significant constraints on the complex scalar DM model parameter space within the region of interest explicitly explored here. If the DP 
mass is larger than 100 MeV and/or $\epsilon$ is constrained to be smaller by other experiments, then these constraints will only become stronger.

\section{Pseudo-Dirac Fermion DM Scenario}

We now turn to the case of a pseudo-Dirac DM particle; here the complex scalar part of $L_{DM}$ above is replaced by 
\begin{equation}
L_{DM}  =i\bar \chi\gamma^\mu D_\mu \chi -m_D \bar \chi \chi -(y_s \bar \chi \chi^c S+h.c.) \,,  
\end{equation}
where, in addition to the $U(1)_D$-invariant Dirac mass, $m_D$, we assume that there also exists a Majorana mass term, $m_M$, 
which is generated by $\chi$'s coupling to $S$ after it obtains the 
dark Higgs vev, $v_s$. For this to happen, of course, we need to require that $2Q_D(\chi)+Q_D(S)=0$ which is easily arranged. In such a case, the Dirac field $\chi$ splits into two distinct 
mass eigenstates:  $\chi=(\chi_1+i\chi_2)/\sqrt {2}i$ and $\chi^c=i(\chi_1-i\chi_2)/\sqrt {2}$,  with $m_{1,2}=m_D\mp m_M$ where $m_M=y_sv_s/\sqrt{2}$ and $y_s$ being a (assumed real) 
Yukawa coupling{\footnote {Note that in our notation $\chi_1$ is the lighter state and we will for simplicity assume that $m_M<m_D$ here.}}. The fractional mass splitting between these 
two states is then simply $\delta=(m_2-m_1)/m_1=2m_{M}/m_1$ which may, in principle, be $O(1)$ or even larger. For values of $\delta \gsim 0.01-0.05$ or so, it is unlikely that any 
signal would be obtained in this scenario from direct detection experiments\cite{TuckerSmith:2001hy} at tree level since the DM trapped in the galaxy would have insufficient velocity to 
excite the higher mass state.  Here $\chi_1$ is identified with the stable DM whereas $\chi_2$ can now decay to $\chi_1$ plus a (possibly on-shell) DP due to an off-diagonal interaction. 
For smaller values of $\delta$, where $V$ is clearly off-shell as in the cases we consider here, this lifetime can be fairly long since the decay width roughly scales as 
$\sim (g_De\epsilon)^2\delta^5$\cite{DeSimone:2010tf,Krall:2017xij}. In the mass eigenstate basis, this leading order off-diagonal interaction of $\chi_{1,2}$ with the DP, $V$, is given by 
\begin{equation}
L_{int}  =ig_D\Big( \bar \chi_1 \gamma_\mu \chi_2-\bar \chi_2 \gamma_\mu \chi_1 \big)V^\mu\,, 
\end{equation}
from which we see immediately why co-annihilation via virtual $s$-channel $V-$exchange is necessary to obtain the observed DM relic density since the `direct' reactions 
$\bar \chi_{1(2)} \chi_{1(2)} \to V^* \to e^+e^-$ do not occur at lowest order in $g_D$. 

As is well-known\cite{Griest:1990kh}, the co-annihilation process $s$-channel cross section for $\bar \chi_1\chi_2 +h.c.\to V^*\to e^+e^-$ 
at freeze-out is suppressed by a factor of $\lambda=2F/(1+F)^2$\cite{Griest:1990kh}, where $F=(1+\delta)^{3/2}e^{-\delta x_f}$, in comparison to the naive calculation due to the thermally 
suppressed $\chi_2$ distribution. For $\delta=0.1(0.2,0.3)$ one finds that $F\simeq 0.16(0.024,0.0037)$, assuming that $x_f=20$, and falls quite rapidly as $\delta$ increases further 
-- hence the reason why small values of $\delta$ are clearly preferred in obtaining sufficiently large cross sections. In the non-relativistic limit for this $s$-wave process we find 
that this annihilation cross section, using the notation above (but now with $r=m_1/m_V$), is given by 
\begin{equation}
\sigma_{DM} v_{rel}=34.1~\lambda \sigma_0 ~\Big(\frac{g_D\epsilon}{10^{-4}}\Big)^2 ~\Big(\frac{100 \rm {MeV}}{m_V}\Big)^2~ \frac{r^2(1+\delta)}{[r^2(2+\delta)^2-1]^2+G^2}\,.  
\end{equation}
Here we have assumed that the contributions to $<\sigma_{DM} v_{rel}>$ from both the $\bar \chi_1 \chi_1$ and $\bar \chi_2 \chi_2$ channels can be neglected since they appear only at 
higher order in $g_D^2$ (as will be discussed below).

At later times, \eg, during the CMB (at roughly $z\sim 10^3$), this co-annihilation cross section is suppressed by a factor of order $\sim e^{-\delta x_f (T_{FO}/T_{CMB})}$ as previously noted 
so that the CMB constraints are simultaneously very easily satisfied. We note that if $m_1>m_V$ then the $s$-wave cross section for $\bar \chi_1 \chi_1 \to 2V$ with $V$ on-shell is 
no longer $\epsilon^2$ suppressed and can easily lead to conflict with CMB constraints as was the case for $\phi^\dagger \phi \to 2V$ above; thus we will maintained the requirement 
that $r=m_1/m_V<1$ in our calculations here to avoid this issue.

To get a feel for these cross sections we will assume as above that $m_V=100$ MeV and $g_D\epsilon=10^{-4}$ and obtain results as a function of $r$ for various values of the mass 
splitting $\delta$; for other input parameter choices a simple rescaling employing the equations above is straightforward, once annihilations to additional possible SM final states are 
appropriately accounted for in the cross section as in the case of complex scalar DM. The calculation is performed as in the complex scalar DM model except that the initial state 
distributions are now Fermi-Dirac. The output of this calculation is seen in the top panel of 
Fig.~\ref{fig0}. The falling behavior of the cross section with increasing $r$ is similar to that obtained in the case of complex scalar DM, but we also see that the cross section declines 
significantly as expected as the fractional mass splitting, $\delta$, is also increased for fixed $r$. 
One might expect that as the value of either of these parameters is increased, the value of $g_D\epsilon$ 
required to obtain the observed DM relic density must correspondingly increase and this is indeed the case as shown in the lower panel of this same Figure. Thus we expect that the 
constraints arising from comparisons of dark ISR cross sections with the CMB bounds to be strongest for larger values of either $r$ or $\delta$. Here we also see that the relevant 
required range of values for the product $g_D\epsilon$ is somewhat greater than in the case of complex scalar DM; this is particularly so when $\delta=0.25$ is assumed. As 
in the complex scalar case, for any assumed value of $\epsilon$ this result also tells us the required value of $g_D$. It is to be noted that for smaller values of $\delta$ and fixed values of $r$, 
these annihilation cross sections are larger than those for complex scalar DM so that it is likely that the resulting constraints from the CMB may also be slightly weaker for small values 
of $\delta$ than in that case. 

\begin{figure}[htbp]
\vspace*{-0.5cm}
\centerline{\includegraphics[width=5.0in]{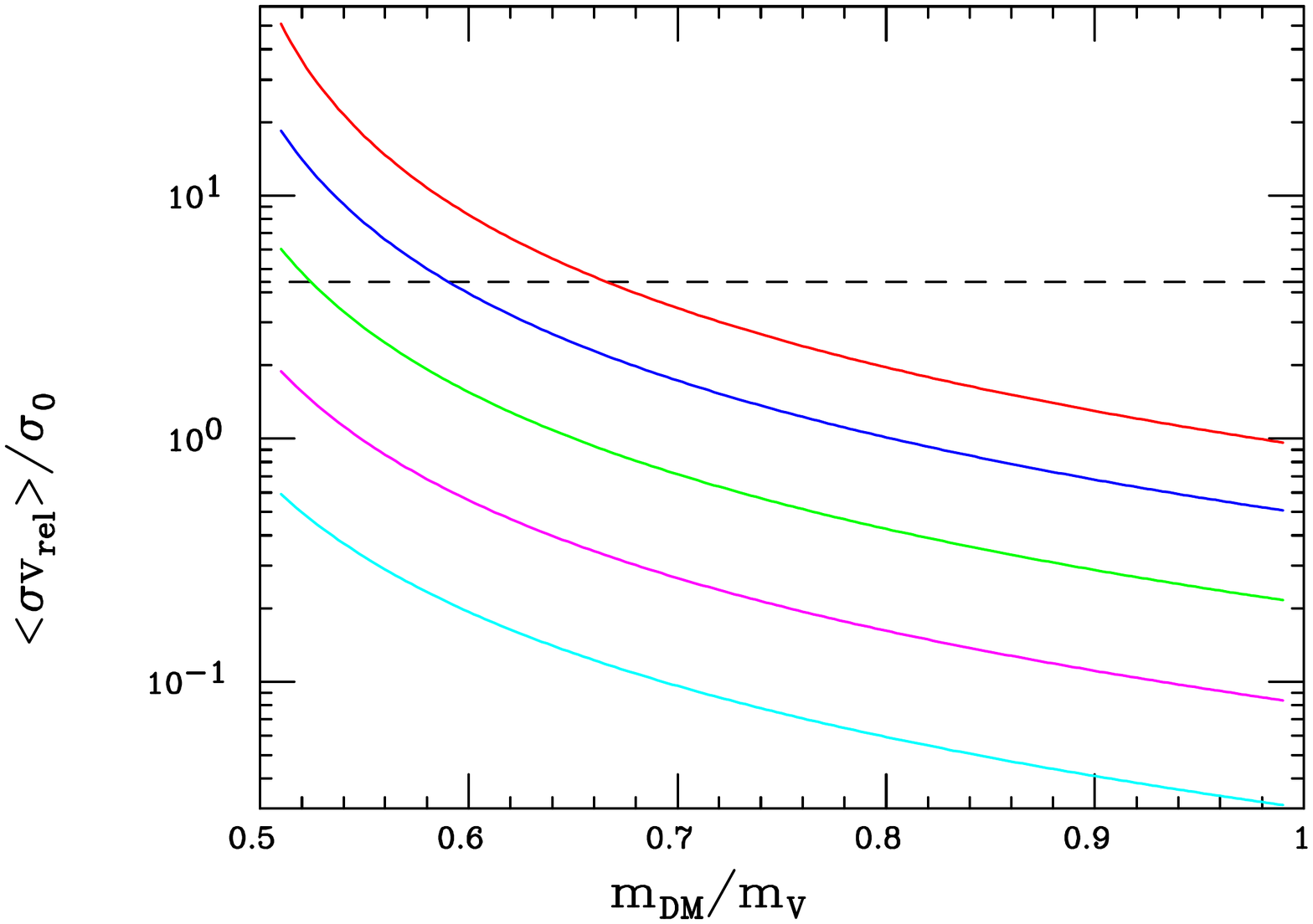}}
\vspace*{-2.2cm}
\centerline{\includegraphics[width=5.0in,angle=0]{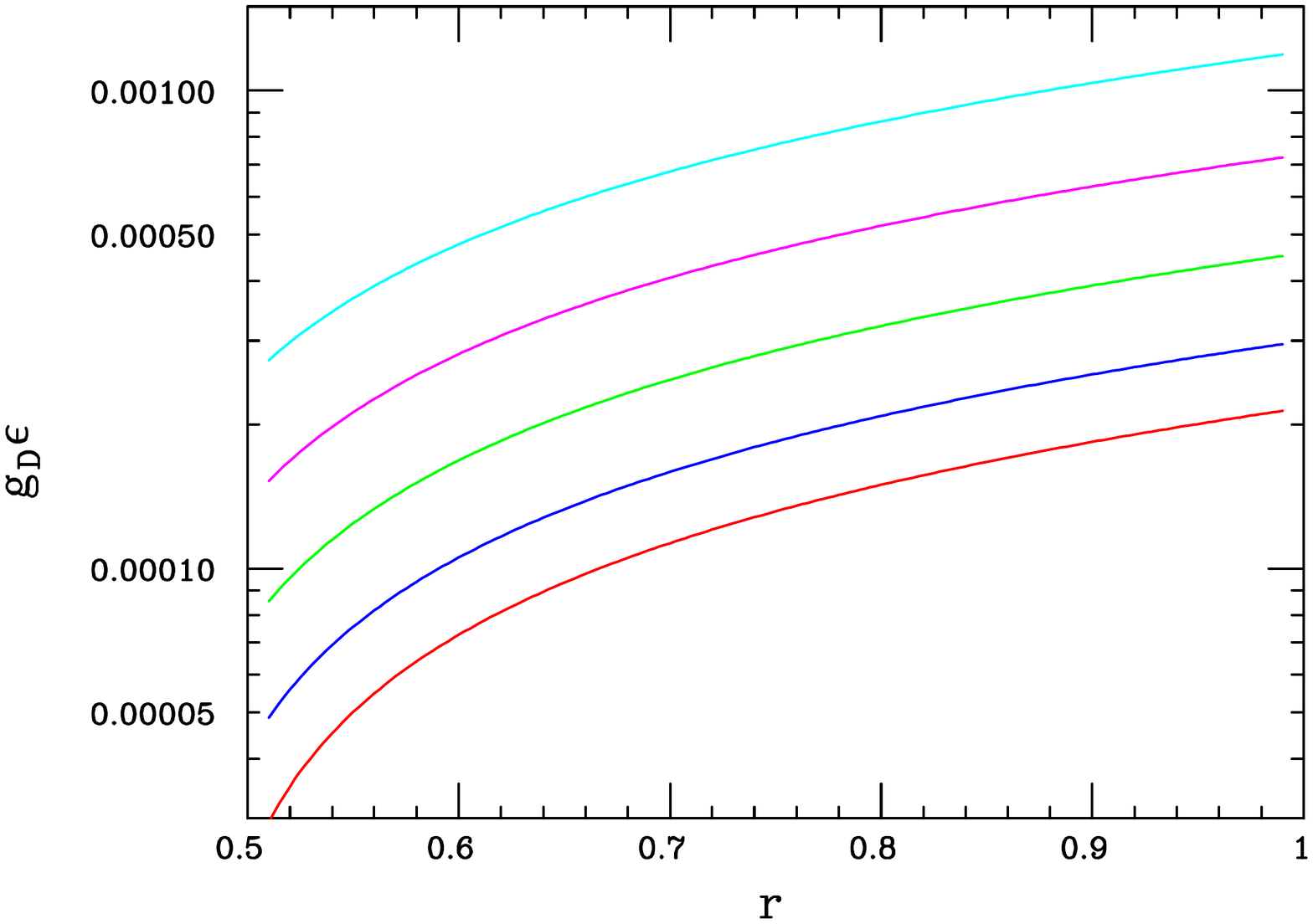}}
\vspace*{-1.50cm}
\caption{(Top) DM co-annihilation cross section in units of $\sigma_0$ assuming that $g_D\epsilon=10^{-4}$ and $m_V=100$ MeV for purposes of demonstration as a function of 
$r=m_{DM}/m_V$. From 
top to bottom the curves correspond to the choices of $\delta=0.05,0.10,0.15,0.20$ and $0.25$, respectively. The dashed line represents the approximate cross section needed to obtain 
the observed relic density. Results for other values of the parameters can be obtained by simple rescaling using the results in the text. (Bottom) Values of $g_D\epsilon$ required to 
obtain the observed relic density employing the same parameter values with curves labelled in reverse order from the top panel.}
\label{fig0}
\end{figure}

With DP ISR, the $s$-wave process $\bar \chi_1 \chi_1 \to VV^*, V^*\to e^+e^-$ becomes possible when $1/2 < r <1$ where the $\chi_2$ is now exchanged in the $t-$ and 
$u-$channels as in Fig.~\ref{fig12}; the corresponding process with $\chi_1$ interchanged with $\chi_2$ still remains doubly Boltzmann suppressed. In this case one may worry that 
the Majorana mass term above now allows for a potentially 
large coupling $\sim m_M/v_s$ of $\bar \chi_1 \chi_1$ to $S$ so that $S$-exchange itself could directly mediate the $\bar \chi_1 \chi_1\to VV^*$ process. While true, it is easy to convince 
oneself that this process is necessarily dominantly $p$-wave and so that is suppressed by $v_{rel}^2\sim T$ at the later, CMB times of relevance here. 

Again employing the non-relativistic limit, we can obtain the $\bar \chi_1 \chi_1 \to VV^*, V^*\to e^+e^-$ annihilation rate which can be written in a manner identical to that obtained for the 
complex scalar DM case above and is given by
\begin{equation}
\sigma_{DM} v_{rel} =0.216~\sigma_0 ~g_D^2 ~\Big(\frac{g_D\epsilon}{10^{-4}}\Big)^2 ~\Big(\frac{100 \rm {MeV}}{m_V}\Big)^2~I\,,  
\end{equation}
where, employing the abbreviation $q=1/(2r)=m_V/2m_1$ as above,  the phase space integral is just $I=2q^2J$ with $J$ being explicitly given by 
\begin{equation}
J=\int_0^{(1-q)^2} dx_{12}\int_{min}^{max} dx_{23}~  \frac{(1-x_{12}+q^2)[(1-x_{13})(x_{13}-q^2)+(1\to 2)]-(x_{13}-q^2)(x_{23}-q^2)(1+q^2)}{[(x_{12}-q^2)^2+(Gq^2)^2][1-x_{12}-q^2+(z^2-1)/2]^2}\,,  
\end{equation}
where we follow the same notation as above and also define the quantities $z=m_2/m_1=1+\delta$ and $x_{13} =1+q^2-x_{12}-x_{23}=1-E_{e^+}/m_1$. Fig.~\ref{fig3} shows the 
numerical results for this phase space integral $I$ as a function of $r$ for various assumed values of $\delta$. As in the case of complex scalar DM, $I$ shows a very strong 
$r$-dependence but whose overall behavior and shape is not much influenced by the different assumed values of $\delta$ in the range we have examined. Of course, as $\delta$ increases 
for fixed $r$, $I$ does indeed decrease by up to factors of a few, but for the range of values shown this effect is rather slight in comparison to the strong observed $r$-dependence that we 
obtain here. 

\begin{figure}[htbp]
\centerline{\includegraphics[width=5.0in,angle=0]{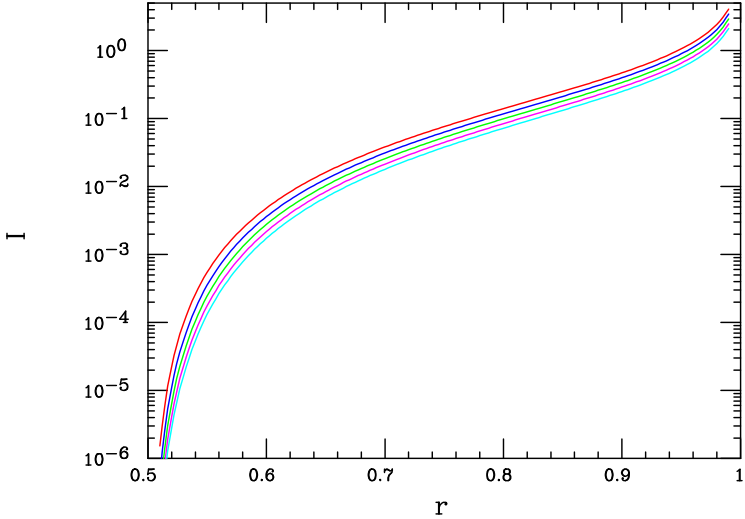}}
\vspace*{-1.30cm}
\caption{ The phase space integral $I$ as defined in the text as a function of $r$ for the case of psuedo-Dirac DM; from top to bottom the curves correspond to the choices 
of $\delta=0.05,0.10,0.15,0.20$ and $0.25$, respectively, as discussed in the text.}
\label{fig3}
\end{figure}

We now proceed as we did earlier in the case of complex scalar DM. We assume that $m_V=100$ MeV for purposes of demonstration and then input the values of $g_D\epsilon$ 
that are required to obtain the observed relic density as a function of $r$ as found above for different assumed values of $\delta$. We then assume as input 
the same three representative values for $\epsilon=10^{-3},~10^{-3.5}$ and $10^{-4}$, respectively, as we did before from which $g_D$ can then be obtained as a function of both 
$r$ and $\delta$. As in the case of complex scalar DM we will always subject the value of $g_D$ to the perturbativity constraint as employed above, \ie, $g_D\leq \sqrt {4\pi} \simeq 3.55$.
As in the case of complex scalar DM, we note that along each curve of constant $\epsilon,\delta$, the value of 
$g_D$ will increase from left to right along the curve as $r$ increases to maintain the value of the 
observed relic density and whose value can be extracted from the lower panel of Fig.~\ref{fig0}.

\begin{figure}[htbp]
\vspace*{-0.5cm}
\centerline{\includegraphics[width=5.0in]{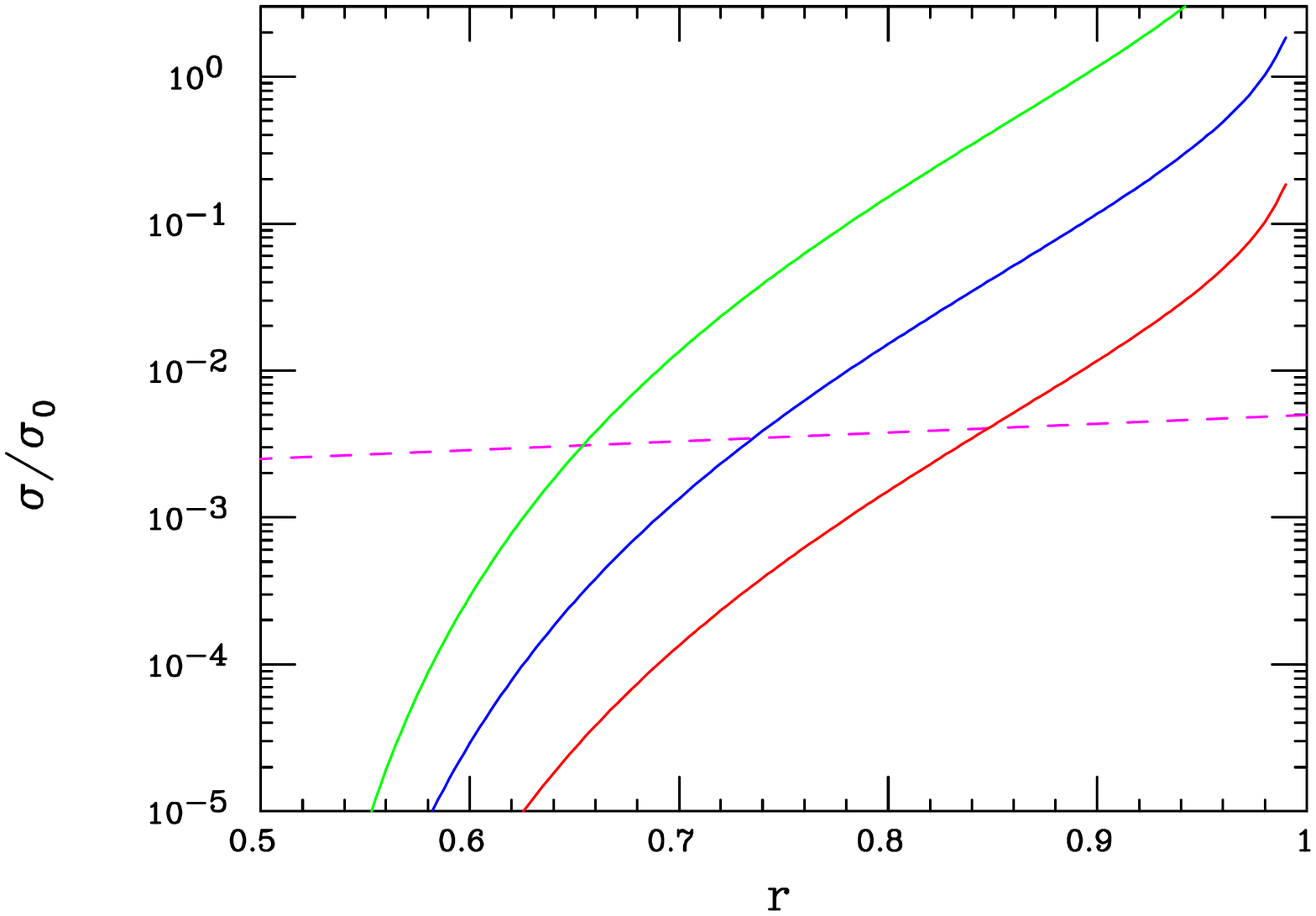}}
\vspace*{-2.2cm}
\centerline{\includegraphics[width=5.0in]{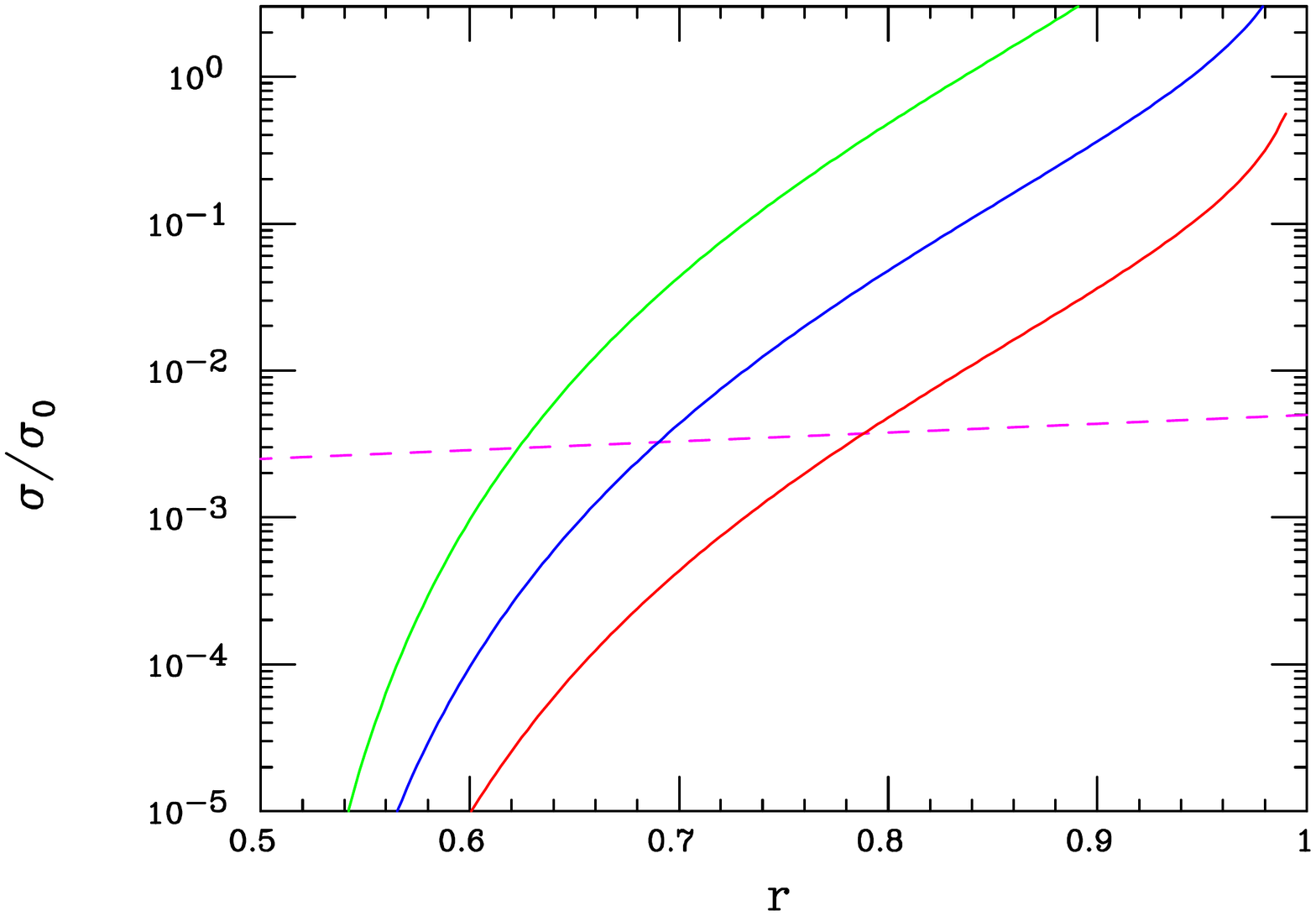}}
\vspace*{-1.50cm}
\caption{(Top) $s-$wave 3-body DP ISR annihilation cross section as a function of $r$ in units of $\sigma_0$ assuming $\delta=0.05$ and $m_V=100$ MeV with, from bottom to top, 
$\epsilon=10^{-(3,3.5,4)}$. The approximate upper limit on this cross section arising from the CMB\cite{Cang:2020exa} is also shown as the dashed line. (Bottom) Same as the top panel 
but now with $\delta=0.1$.}
\label{fig7}
\end{figure}

The results of this calculation for the assumed range values of $\delta=0.05-0.25$ can be found in the set of Figs.~\ref{fig7},~\ref{fig8} and ~\ref{fig9} which show the cross section 
for the process $\bar \chi_1\chi_1 \to VV^*, V^*\to e^+e^-$ with $v_{rel}^2=0$ in units of $\sigma_0$ as a function of $r$ for the three values of $\epsilon$ chosen above 
taking $m_V=100$ MeV.  Also shown is the 
approximate constraint arising from the CMB\cite{Cang:2020exa}. In all 5 of these cases the overall qualitative nature of the results are similar to what was found for the case of complex 
scalar DM but the details are observed to be quite different. For fixed values of $\delta$ and $\epsilon$ the cross section rises quickly with increasing $r$ due to both the rapid opening of 
the phase space as well as the increasing value of $g_D$ as previously observed. As an example, for a fixed value of $\delta=0.05[0.1]$, assuming that $\epsilon=10^{-3},~10^{-3.5}$ and 
$10^{-4}$ leads to the constraint $r \lsim 0.85,0.74,0.65[0.79,069.0.62]$, respectively. 
However, as $\delta$ increases for fixed $\epsilon$ we also see that the bound arising from the CMB rapidly 
becomes stronger as might be expected since for fixed values of $r$ and $\epsilon$ the value of $g_D$ must increase to obtain the observed relic density. For example, assuming 
that $\epsilon=10^{-3.5}$ we obtain an upper limit of $r \lsim 0.74$(0.69,0.64,0.60,0.57) for $\delta=0.05$(0.1,0.15,0.2,0.25), respectively. One difference we see with the earlier result 
for the complex scalar case is that when $\delta=0.2,0.25$ and $\epsilon=10^{-4}$ the predicted cross section curve terminates in the middle of the plot but still within the now 
excluded region. This is 
the result of imposing the perturbativity constraint on $g_D$ as discussed above, \ie, $g_D\leq \sqrt {4\pi} \simeq 3.55$, \ie, larger values of $r$ in these case requires values of 
$g_D$ exceeding the perturbativity bound in order to reach the desired relic density. These values of $g_D$ can be read off directly by using the lower panel of Fig.~\ref{fig0} for any 
given value of $\epsilon$. As in the scalar DM case 
above, we again should notice in the Figures is that for larger values of $r$ the 3-body process could have made a genuine contribution to the DM 
relic density if this region had not already been disallowed by the CMB data.

\begin{figure}[htbp]
\vspace*{-0.5cm}
\centerline{\includegraphics[width=5.0in]{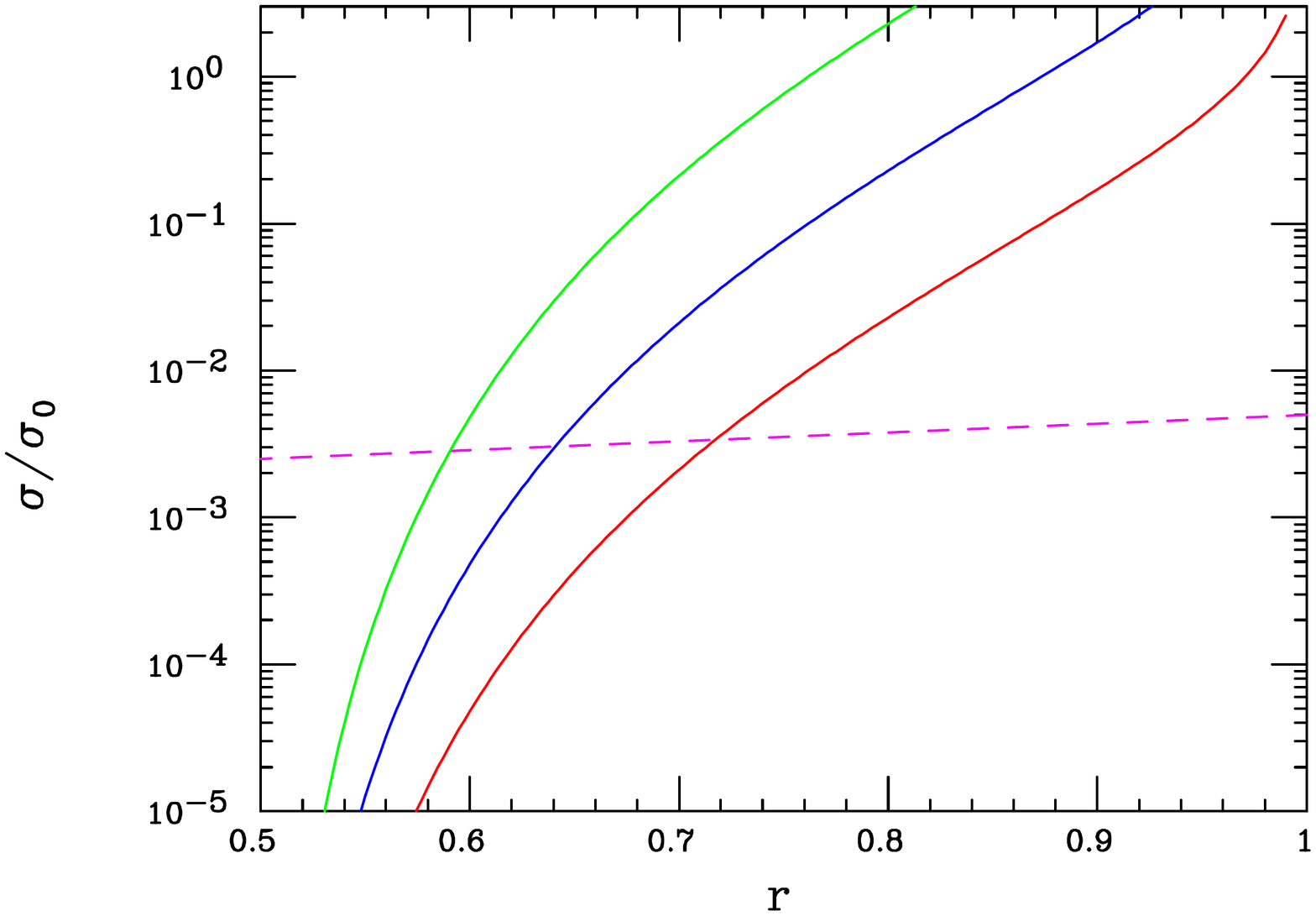}}
\vspace*{-2.2cm}
\centerline{\includegraphics[width=5.0in]{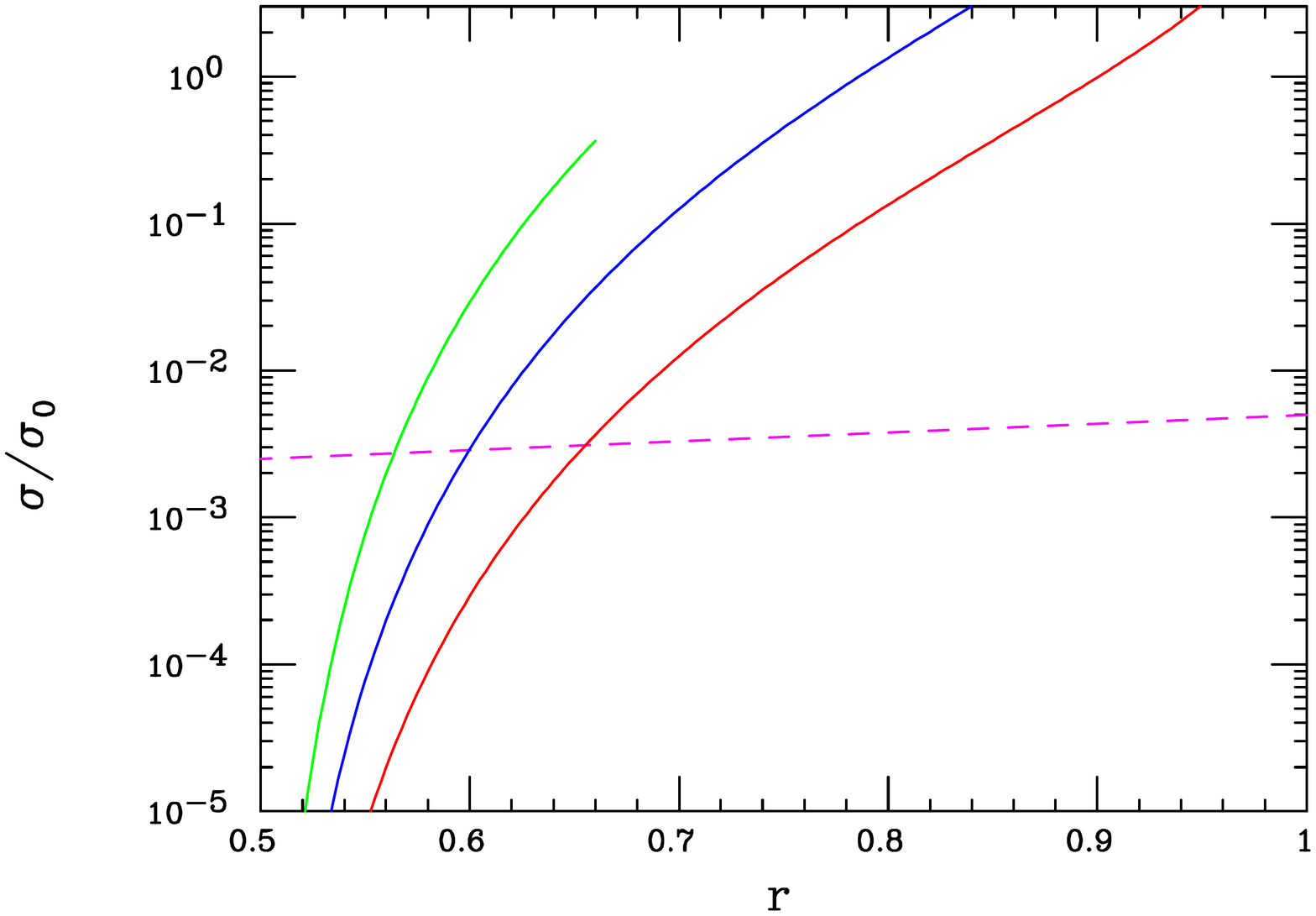}}
\vspace*{-1.50cm}
\caption{Same as the previous Figure but now assuming that $\delta=0.15$ (Top) or $\delta=0.2$ (Bottom). Note the termination of the curve for $\epsilon=10^{-4}$ in the middle of the 
plot due to the perturbativity constraint on $g_D$.}
\label{fig8}
\end{figure}

As was observed in the previously examined case of complex scalar DM, it is clear from this analysis that the CMB can place significant constraints on the pseudo-Dirac DM model parameter 
space. If the DP mass is larger than 100 MeV, $\epsilon$ is constrained to take on values smaller than those considered here by other experiments, or the mass splitting between the 
two fermion states is for some reason larger, then in any of these cases the constraints shown above will only become stronger. Clearly this is an important constraint on this model 
in this kinematic range.

\begin{figure}[htbp]
\vspace*{-0.5cm}
\centerline{\includegraphics[width=5.0in]{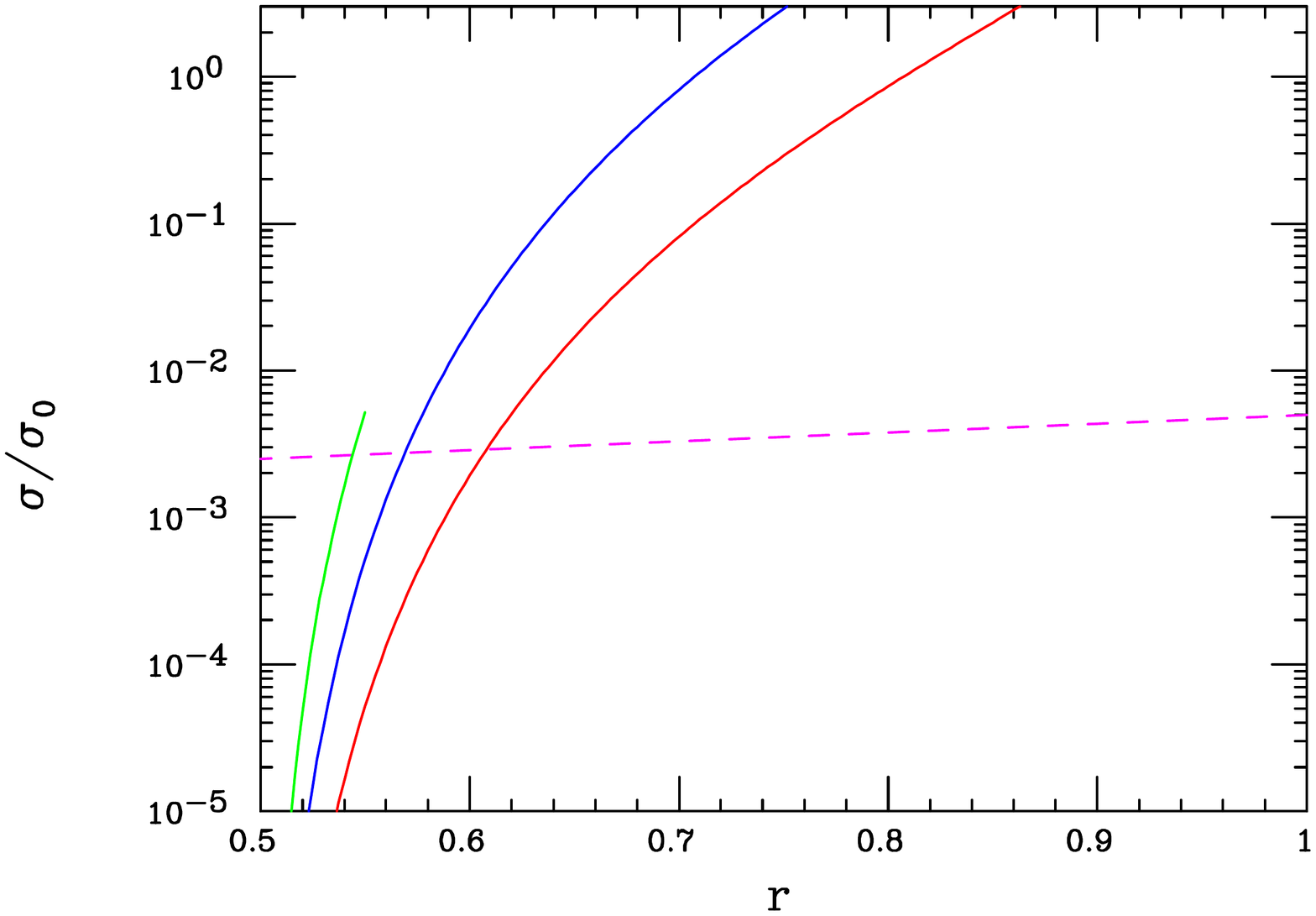}}
\vspace*{-1.50cm}
\caption{Same as the previous Figure but now assuming that $\delta=0.25$. Again the perturbativity constraint on $g_D$ is playing a role here when $\epsilon=10^{-4}$. }
\label{fig9}
\end{figure}

\vspace{1cm}
\section{Discussion and Summary}

The kinetic mixing model with both dark photons and dark matter in the sub-GeV mass range poses an interesting alternative to the well-studied traditional WIMP and axion scenarios. 
However, constraints from the CMB can pose significant model building requirements on this class of models in that the annihilation process by which the DM achieves the observed 
relic density cannot be dominantly $s$-wave, or more generally, temperature-independent. Annihilation via a $p-$wave process, as in the case of complex scalar DM, or via co-annihilation, 
as in the case of pseudo-Dirac DM with a small mass splitting, offer two attractive alternative setups that circumvent these constraints. In both these scenarios, the emission of additional dark 
ISR in the form of a DP as part of the annihilation process, though higher order in $g_D$ and a 3-body process, 
necessarily leads to a numerically suppressed $s-$wave process with a rate that is 
insufficient to explain the DM relic density but may still be in conflict with the bounds from the CMB in certain parameter space regimes. This leads to an additional set of constraints on 
the model space of these scenarios and, in this paper, we performed a preliminary examination of the impact of these constraints arising from the rate for the emission of this additional DP. 
Indeed, potentially important constraints were obtained on the parameter spaces for both the complex scalar and pseudo-Dirac scenarios once the necessity of achieving the observed 
DM relic density was simultaneously imposed. 

For both of these DM models we followed the same procedure: fixing the DP mass to 100 MeV, we first determined the value of the product $g_D\epsilon$ as a function on the DM to 
DP mass ratio, $r$, required to obtain the observed DM relic density. In the pseudo-Dirac case this analysis was performed for several values of the mass splitting between the two 
states, $\delta$. Using scaling and the measured value of $R(e^+e^- \to hadrons)$, the corresponding result can be obtained for any value of $m_V$; larger values of $m_V$ generally 
require larger values of $g_D\epsilon$ for fixed $r$. These results were then employed as input into the dark ISR rate calculation for several fixed values of $\epsilon$ and results for any 
other set of  
values of $\epsilon$ are easily obtainable from simple scaling. This fixed the numerical value of $g_D$ and only those parameter space points where $g_D$ remained perturbative, \ie, 
$g_D \leq \sqrt {4\pi}\simeq 3.55$, were kept. 

For both the complex scalar and pseudo-Dirac DM models with the DP and DM within the ranges considered here, rather strong constraints on the the relevant parameter spaces were 
obtained.  Though qualitatively similar, these constraints differ in detail between these two types of models. In both models, for fixed values of the DP mass, $m_V$, as well as for fixed 
$\delta$ in the pseudo-Dirac DM case, when the KM parameter, 
$\epsilon$ becomes smaller, a correspondingly larger value of the dark gauge coupling, $g_D$, is needed to obtain the necessary DM annihilation cross section for a fixed value of the 
ratio of the DM and DP masses, $1/2<r=m_{DM}/m_{DP}<1$. As this mass ratio increases,  since the cross section for dark ISR scales as $g_D^4$, regions of the parameter with 
larger values of $r$ are found to be more constrained -- and indeed become excluded -- by the CMB measurements. Since $g_D$ must increase when $m_V$ increases, taking all other 
parameters fixed, the bound on $r$ from the CMB only becomes stronger. Similarly, in the pseudo-Dirac DM model, increasing $\delta$ with all other parameters fixed also strengthens 
the bound on $r$ with part of the parameter space being eliminated by the perturbativity requirement on $g_D$.  For smaller values of $\delta \lsim 0.05-0.1$, the bounds obtained on $r$ 
are somewhat weaker in the pseudo-Dirac case in comparison to the complex scalar DM model due to the somewhat larger cross section for co-annihilation. However, for larger values of 
$\delta \gsim 0.2-0.25$ this situation is reversed since the co-annihilation cross section is now more highly Boltzmann suppressed. Furthermore, in both of these models, decreasing 
the value of $\epsilon$ itself while holding everything else fixed raises the dark ISR cross section as it scales as $\epsilon^{-2}$. Again, this will lead to even stronger constraints on the 
other model parameters.

From the above analysis it is clear that further detailed study of the effects of the CMB constraints and dark ISR on the parameter spaces of light DM models is warranted.

\section {Acknowledgements}
The author would like to thank J.L. Hewett and G. Wojcik for valuable discussions related to this analysis, Yu-Dai Tsai for both discussions and various numerical checks,   
and also D. Rueter for both discussions and for providing the Feynman diagrams used within.  The author would also like to thank T. Slatyer for bringing the 
work in Refs.\cite{Boudaud:2016mos,Boudaud:2018oya} to his attention. This work was supported by the Department of Energy, Contract DE-AC02-76SF00515.



\end{document}